\documentclass[conference]{IEEEtran}
\usepackage{cite}
\usepackage{amsmath,amssymb,amsfonts}
\usepackage{algorithmic}
\usepackage{graphicx}
\usepackage{textcomp}
\usepackage{xcolor}

\usepackage{multirow}

\def\BibTeX{{\rm B\kern-.05em{\sc i\kern-.025em b}\kern-.08em
    T\kern-.1667em\lower.7ex\hbox{E}\kern-.125emX}}
\begin{document}

\title{Crowd-sensing commuting patterns using multi-source wireless data: a case of Helsinki commuter trains\\
}

\author{\IEEEauthorblockN{1\textsuperscript{st} Zhiren Huang}
\IEEEauthorblockA{\textit{Dept. of Computer Science} \\
\textit{Aalto University}\\
Espoo, Finland \\
0000-0002-7868-1630}
\and
\IEEEauthorblockN{2\textsuperscript{nd} Alonso Espinosa Mireles de Villafranca}
\IEEEauthorblockA{\textit{Dept. of Built Environment} \\
\textit{Aalto University}\\
Espoo, Finland \\
0000-0002-4521-0234}
\and
\IEEEauthorblockN{3\textsuperscript{rd} Charalampos Sipetas}
\IEEEauthorblockA{\textit{Dept. of Built Environment} \\
\textit{Aalto University}\\
Espoo, Finland \\
0000-0002-9829-3483}
\and
\IEEEauthorblockN{4\textsuperscript{th} Tri Quach}
\IEEEauthorblockA{\textit{Dept. of Analytics and Research} \\
\textit{HSL Helsingin seudun liikenne}\\
Helsinki, Finland \\
}
}

\maketitle

\begin{abstract}
Understanding the mobility patterns of commuter train passengers is crucial for developing efficient and sustainable transportation systems in urban areas. Traditional technologies, such as Automated Passenger Counters (APC) can measure the aggregated numbers of passengers entering and exiting trains, however, they do not provide detailed information nor passenger movements beyond the train itself. To overcome this limitation we investigate the potential combination of traditional APC with an emerging source capable of collecting detailed mobility demand data. This new data source derives from the pilot project TravelSense, led by the Helsinki Regional Transport Authority (HSL), which utilizes Bluetooth beacons and HSL's mobile phone ticket application to track anonymous passenger multimodal trajectories from origin to destination. By combining TravelSense data with APC we are able to better understand the structure of train users' journeys by identifying the origin and destination locations, modes of transport used to access commuter train stations, and boarding and alighting numbers at each station. These insights can assist public transport planning decisions and ultimately help to contribute to the goal of sustainable cities and communities by promoting the use of seamless and environmentally friendly transportation options.
\end{abstract}

\begin{IEEEkeywords}
mobile application, Bluetooth beacon, automated passenger counting, public transport, multimodal
\end{IEEEkeywords}

\section{Introduction}

In today's urban areas, public transportation (PT) plays a crucial role in achieving sustainability goals by reducing car usage and emissions \cite{Newman2015, Ceder2021}. Commuter trains are an important mode of transportation for many people, especially in urban areas where distance, traffic congestion, and parking can be a problem \cite{Wu2019}. Understanding the mobility patterns of commuter train passengers is therefore crucial for developing efficient and sustainable transportation systems. Commuter train ridership \cite{Lin2016, Borjesson2019, Jenelius2020, Sipetas2020}, travel time and distance distributions \cite{Zhao2015, Zhang2016}, access and egress modes \cite{Cervero2001, Bhat2006, Givoni2007, Chakour2013, Jonkeren2019, Giansoldati2021}, and the origin and destination of commuter train trips \cite{Halldorsdottir2017} are all important factors in understanding how people move within a city \cite{Cai2022}. However, traditional methods for collecting this data, such as travel surveys \cite{Raty2018}, are not only time-consuming and costly but are also performed yearly or even more infrequently. This makes it difficult to capture real-time or near-real-time data. While Automated Passenger Counters (APC) can provide aggregated information on the number of passengers entering and exiting trains \cite{Nassir2011, Roncoli2023}, they do not provide detailed information nor passenger movements beyond the train system.  In addition, high costs of installing such devices often leads to low penetration rates within PT systems. These limitations make it difficult for transportation planners and managers to make informed decisions about PT systems.

To overcome these limitations, we investigate the potential combination of conventional APC data with an emerging source of data that can provide a more detailed understanding of passenger mobility. The new data source derives from the pilot project TravelSense \cite{Huang2022}, led by the Helsinki Regional Transport Authority (HSL). The project utilizes Bluetooth beacons and HSL's mobile phone ticket application to track anonymous passenger multimodal trajectories from origin to destination. 

Since this type of door-to-door trajectory can be highly sensitive with regards to privacy, several methods are applied in collecting and in pre-processing the data in order to preserve enough anonymity to avoid identification of individuals: 
\begin{itemize}
    \item The data collected derives from users who have explicitly opted-in to share their mobility data with HSL. This means that users need to turn on the travel data sharing option in the settings by themselves.
    \item Each user's mobile phone is assigned a random ID for every day about which the user shares data. This prevents tracking the same individual over several days.
    \item Spatial resolution for actual origins and destinations are coarse grained and timestamps obfuscated for recognitions outside the PT network.
\end{itemize}

The challenge that arises is captured by the question: does the sparseness of the data significantly affect its usefulness while also restricting its applications? We compare the ridership deriving from TravelSense with the one from APC to validate the TravelSense data representativeness. The assumption is that if the comparison of ridership between the two data sources shows high correlation, then other travel patterns sensed from TravelSense (e.g., travel time and distance distributions, access and egress modes, the origin and destination of commuter train trips) are also more reliable than what the above-mentioned concerns might imply.

By combining TravelSense data with APC data, we are able to better describe and understand the commuting patterns of train users. These insights can assist PT planning decisions and ultimately contribute to the sustainability goals of cities and communities by promoting the use of environmentally friendly transportation options and by helping to enable seamless and user-centric transport systems. The main contributions of this study are:
\begin{itemize}
    \item Utilization of the new data source, TravelSense, to demonstrate how mobile data technologies can enhance the power of sensing near-real-time PT states.
    \item Development of a data integration framework to capture the commuting train mobility patterns for a metropolitan region.
\end{itemize}

The paper is structured in two parts. First, we compare the representativeness of the TravelSense dataset with commuting train APC data; second, we combine the two sources to understand the commuting patterns of train passengers.

\section{Study area and data}

\begin{figure}[tb]
\centerline
{\includegraphics[width=9cm]{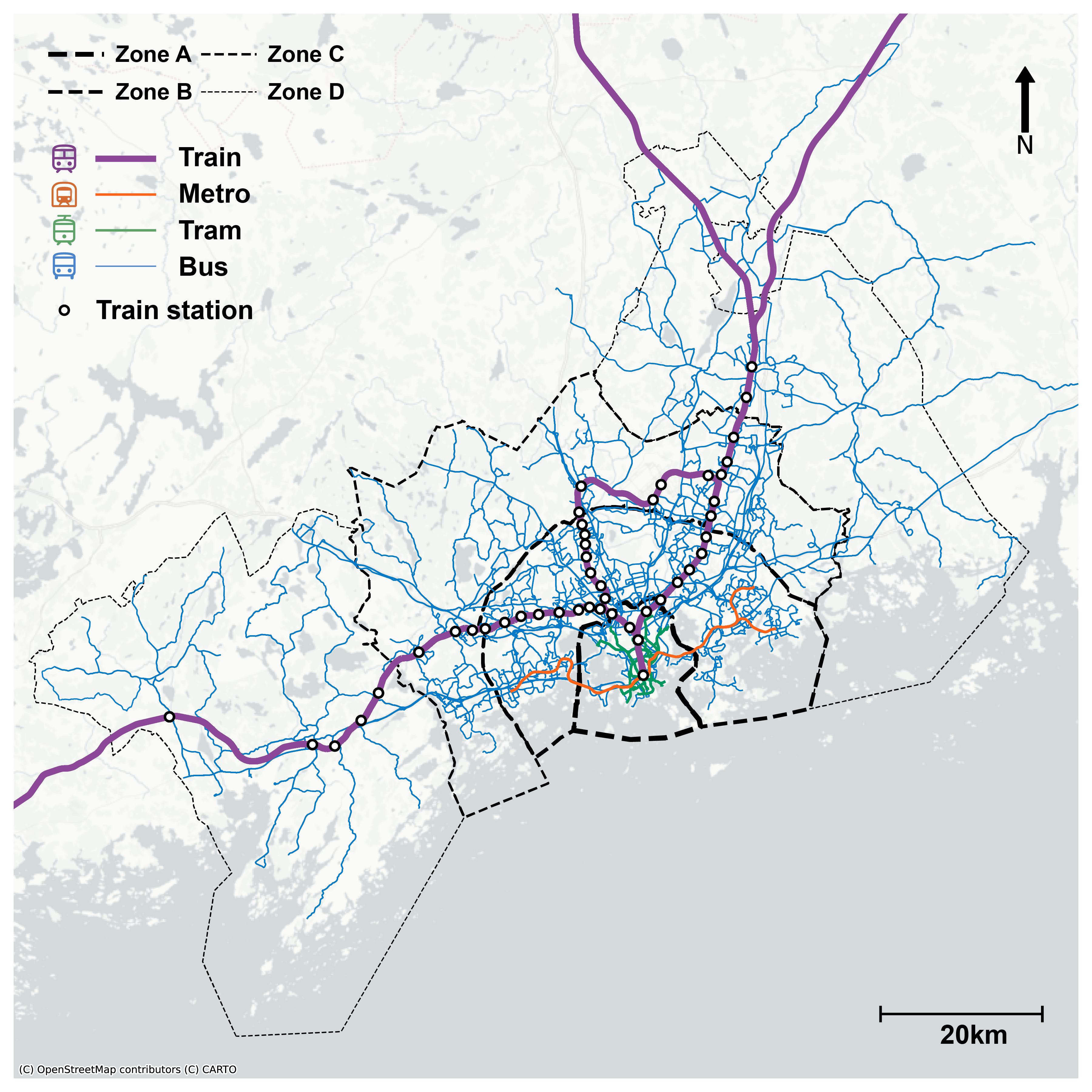}}
\caption{The study area and the PT network including different transportation modes: train (purple), subway (orange), tram (green), and bus (blue). Dash lines represent four ticket zones based on thickness.} 
\label{fig:train_map}
\end{figure}

\subsection{Study area}

The Helsinki metropolitan area is approximately 770 km$^2$ with a population of approximately 1.2 million inhabitants. The local mobility services include fixed PT (metro, tram, train, bus, and ferry), micro-mobility (shared e-scooters and shared bicycles), and ride-hailing services (e.g., UBER), which vary according to location. A map of the study area including the PT network is presented in Fig.~\ref{fig:train_map}. Purple lines represent the commuter train network.

HSL provides PT services for 9 municipalities (HSL area) which, in addition to the capital region, include Siuntio, Kirkkonummi, Sipoo, Kerava, and Tuusula \cite{Raty2018}. To facilitate PT trips HSL offers a mobile phone ticketing application that provides travelers with information on PT operation (e.g., fare zones, real-time PT vehicle locations, timetables, best routes, delays, and PT news, among others). 

\subsection{Data}
\label{data}

The datasets used in this paper (TravelSense and APC counts) were provided by HSL. They correspond to the same collection period from the 1st to the 30th of September 2021; details are given below.

\paragraph{Automated Passenger Counters (APC)}

The APC system of commuter trains records the number of passengers entering and exiting train vehicles in real-time, making use of sensors installed on-board commuter train vehicles. The number of units installed varies from one to three units per vehicle in Helsinki trains. The APC dataset records nine routes and 45 train stations within the Helsinki PT network. It is noted that in rare cases the APC devices do not operate properly due to technical problems, and this leads to the respective trip counts being imputed by the PT operator. The percentage of imputed values is usually low and differs from day to day, since the occurrence of technical problems is random and unpredictable. In this study, the imputed counts, which are based on historical data, are treated equally to real counts. 

\paragraph{TravelSense}

\begin{figure}[tb]
\centerline
{\includegraphics[width=8.5cm]{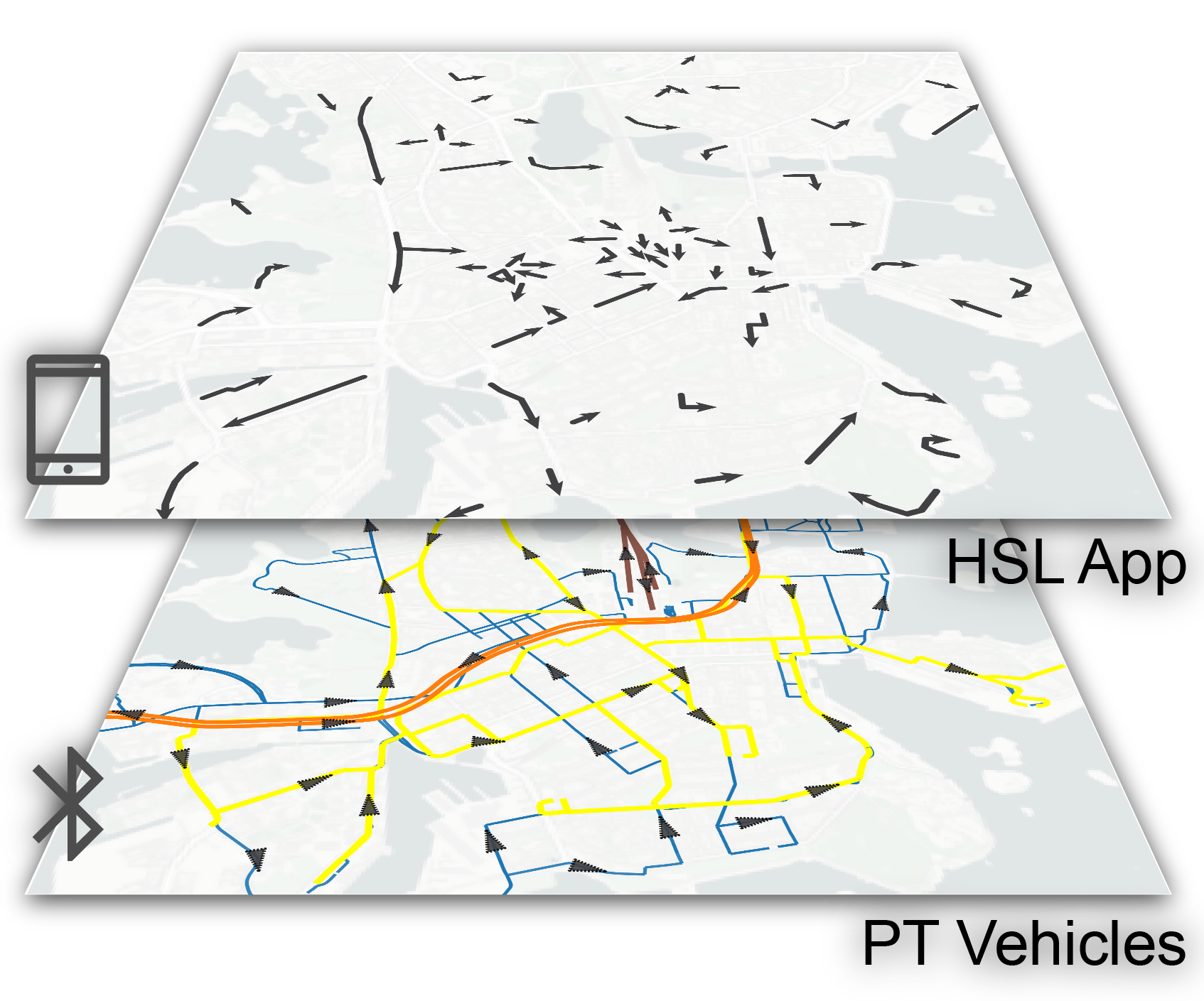}}
\caption{Data collection infrastructure of TravelSense.}
\label{fig:infrastructure}
\end{figure}

The data collection infrastructure principally relies on the HSL mobile ticketing app. Through the app, a user's device is able to recognise the Bluetooth low energy beacons that are present throughout the PT network. The app also uses the mobile phone's GPS coordinates to determine the grid cells through which the user moves. Finally, the app also uses the activity recognition modules of the user's mobile phone to determine whether the user is still, walking, cycling, or on board of a vehicle. To ensure high standards of data privacy, devices are anonymised and given random IDs every day for which data is shared, thus they are not followed longitudinally. Furthermore, the GPS coordinates are resolved only up to grid cells of 250m $\times$ 250m and timestamps of locations outside the PT network are rounded to the nearest quarter-hour.

The physical sources used in the data collection process can be classified in the following way:
\begin{itemize}
    \item Moving Bluetooth beacons. These are installed inside PT vehicles (buses, trams, trains, metros and ferries).
    \item Portable devices. These are the users' mobile phones, which recognise the user activity and movement and also recognise the Bluetooth beacons throughout the PT network.
\end{itemize}
\medskip

The HSL app logs beacon recognition and pushes this information along with position and activity information to the TravelSense servers. The TravelSense data collection system is based on the framework described in \cite{Huang2022} and \cite{Rinne2017}. Currently the data collected is from users who have explicitly opted-in to share their mobility data with HSL; they can choose to stop sharing at any time.

We use a python library \textit{tstrips} developed in our previous study \cite{Huang2022} to process the TravelSense data. The information is structured based on \emph{legs} and \emph{trip chains}, with the following definitions:
\vspace{2mm}
\begin{itemize}
    \item Leg -- an individual segment within a trip chain recognised by the data collecting system and pre-processing as a discrete stage within the journey, either because there are pauses in the movement, or there is a change in recognised activity.
\vspace{2mm}    
    \item Trip chain -- a series of legs that have been recognised by the system in pre-processing as being part of a single journey. The end-points of trip chains are recognised by prolonged periods of remaining in the same location and no significant changes in activity.
\end{itemize}
These definitions are similar to those in the travel survey operated by HSL\cite{Raty2018}, where the trip chain in this paper equals the journey in HSL's travel survey.

\section{Methods}

\subsection{Preliminaries}

\begin{figure}[tb]
\centerline
{\includegraphics[width=7cm]{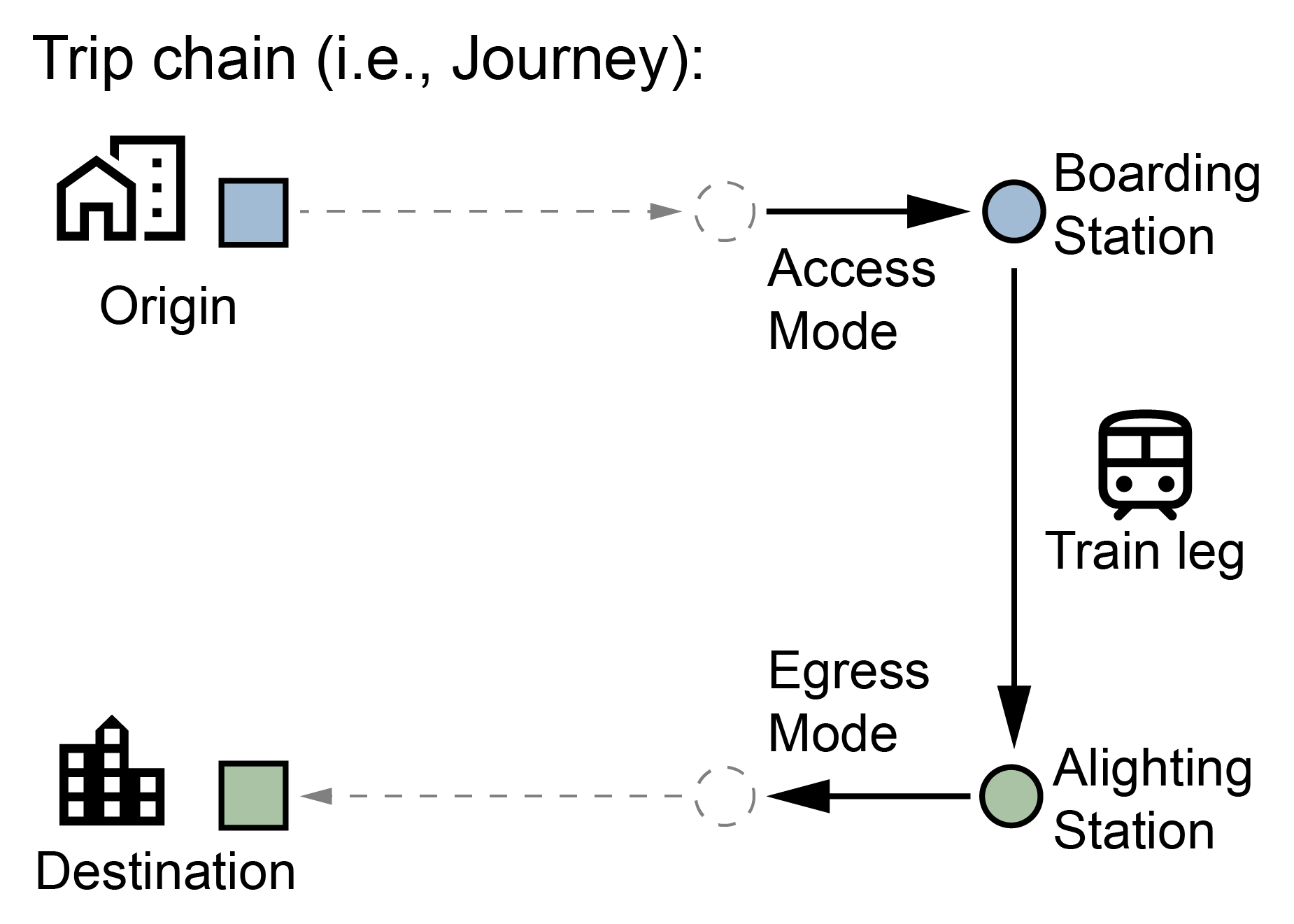}}
\caption{Illustration of the leg and the trip chain.}
\label{illustration}
\end{figure}

Based on the definitions above, a trip chain is characterised by the legs that make up each segment and the travel modes used in each of them. A simplified view of a trip chain is shown in Fig. ~\ref{illustration}, which highlights the main characteristics with which we are concerned.
Namely,
\begin{itemize}
    \item Train leg -- The leg's traffic mode is the commuter train, which contains the timestamp of entering and exiting a train vehicle, as well as the boarding and alighting stations.
    \item Origin \& destination -- The first grid cell of a chain trip which contains a train leg is the origin of this boarding station. The last grid cell of this trip chain is the destination of the train leg’s alighting station.
    \item Access \& egress mode -- The traffic mode of the last leg before the train leg is the access mode of this boarding station. The traffic mode of the leg after this train leg is the egress mode of this alighting station.
\end{itemize}

\begin{figure*}[tb]
\centerline
{\includegraphics[width=15.5cm]{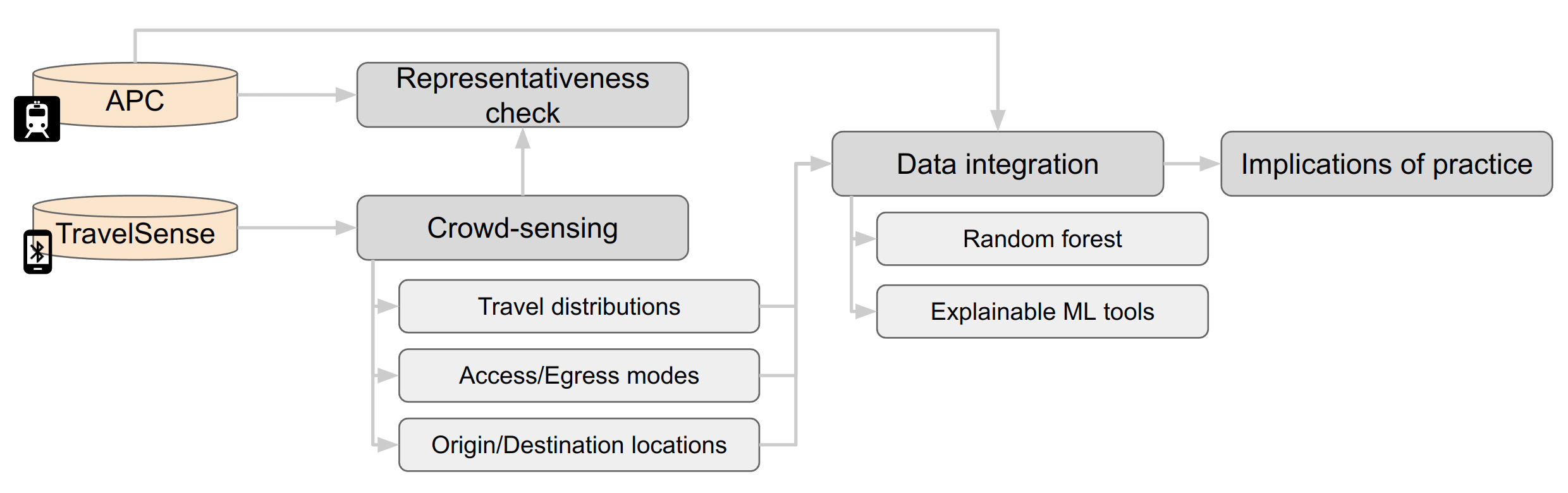}}
\caption{Framework of the proposed methodology.}
\label{framework}
\end{figure*}

\medskip
Fig.~\ref{framework} shows the framework of this study and the arrows show the data flow. Due to the limitation of paper length, in this study, we focus on the data during weekdays. The temporal factors (i.e., hours) are only considered in representativeness check, while crowd-sensing results are presented without temporal factors.

\subsection{Representativeness validation}

In order to determine the representativeness of the TravelSense dataset, we compared it to ridership data collected by APC. APC was chosen as the baseline data source since it is installed on vehicle doors to capture the total number of passengers. APC is shown to be a more reliable data source when compared to other movable devices, such as mobile phones\cite{Roncoli2023} . To quantify the correlation between the two datasets, the following three metrics were used:
\begin{itemize}
    \item Spearman's rank correlation coefficient ($\rho$) -- to measure how well the relationship between the two variables can be described using a monotonic function. $\rho$ ranges from 0 to 1. 
    \item Pearson correlation coefficient (PCC) -- to measure the linear correlation between the two datasets. PCC ranges from -1 to 1. 
    \item R-squared ($R^2$) -- to measure the proportion of the variance of the ridership sensed from APC that is explained by the ridership sensed from TravelSense by a given model. In this study, linear regression was applied. $R^2$ ranges from 0 to 1. 
\end{itemize}

The ridership data was aggregated at the route and station levels; the temporal factor -- the hour of the day was also taken into consideration.

\subsection{Commuting patterns within the commuter train system}
Since APC data only records aggregated data, it is difficult to estimate travel features of a single trip, and therefore it requires sophisticated methods in order to estimate OD matrices of travel demand at the station level \cite{Zhao2007, Hakegard2018}. TravelSense could help in calculating and inferring several important commuting mobility features, such as:
\begin{itemize}
    \item OD matrix -- the number of passengers $T_{ij}$ from the station $i$ to the station $j$ was inferred with the proportion of passengers from the TravelSense whose train leg starts and ends at those respective stops.
    \item Travel time distribution -- the travel time of a train leg was calculated by subtracting the time of boarding from the time of alighting.
    \item Travel distance distribution -- the geographic distance (aka great-circle distance) of a train leg was calculated by the boarding and alighting station coordinates.
    \item Travel flow distribution -- the distribution of travel flow $T_{ij}$ of OD pairs.
\end{itemize}

\subsection{Mobility beyond the train system}

The travel demand serviced by a train station is intrinsically tied to geography and the urban landscape. We use TravelSense to investigate the relationship between passenger loads with spatial (ODs) and behavioural (travel mode) characteristics of journeys that use the train stations.

\paragraph{Origins and destinations identification} \label{sec:OD_id} For each train station, two lists of associated grid cells are compiled corresponding to origins and destinations journeys that use the station. Three metrics are used to quantify the distributions of origins and destinations:

\begin{itemize}
    \item The number of grid cells $N$ -- the number of unique grid cells associated with the train station $k$. $N_{k}^{(o)}$ represents the number of unique origins of a boarding station $k$, $N_{k}^{(d)}$ represents the number of unique destinations of an alighting station $k$.
    \item The Gini index $G$ -- which measures the inequality among the values of a distribution \cite{Dixon1987}. Gini index values close to zero indicate a more uniform distribution, while values closer to one indicate higher inequality. $G^{(o)}$ measures the inequality of the passenger numbers from boarding stations' origins and $G^{(d)}$ measures the inequality of the passenger numbers to alighting stations' destinations.
    \item The radius of gyration $rg$  -- which captures the characteristic distance travelled by a person during a specific time period\cite{Gonzalez2009}. At the population level, the radius of gyration for boarding station's origins is calculated by:
\begin{equation}
    rg_{k}^{(o)} =  \frac{\sum_{i = 1}^{N_{k}^{(o)}} (d_{ik} \times T_{ik})}{\sum_{i = 1}^{N_{k}^{(o)}} T_{ik}}. 
\label{eq_g}
\end{equation}
\noindent  where $d_{ik}$ is the distance between the grid cell $i$ and the station $k$, $T_{ik}$ is the number of passenger from the grid cell $i$ to the station $k$. $rg_{k}^{(d)}$ follows the same form of \ref{eq_g} but for alighting station' destinations.
\end{itemize}
\medskip
\paragraph{Access and egress modes identification}  For each train station, based on the TravelSense trips that use it to access and egress the train network, we assign two vectors which capture the travel modes of access and egress legs. We use eight traffic modes in this study: bus, private car, cycling, subway, train, tram, walking, and unknown. Therefore the length of the access/egress mode vector is eight.

\begin{table*}[htb]
\caption{Correlations between TravelSense and APC}
\label{table_correlations}
\centering
\begin{tabular}{cclllll}
\hline
{Spatial Aggregation} &
  \multicolumn{1}{l}{{Temporal Aggregation}} &
  {Event Type} &
  {Spearman} &
  {PCC} &
  {$R^2$} &
  {Linear Regression} \\ \hline
{} &
  {Weekday} &
  {Boardings/Alightings} &
  {1.000*} &
  {0.991*} &
  {0.982} &
  {40.345$\times x$-10932.415} \\
\multirow{-2}{*}{{Route}} &
  {Hour} &
  {Boardings/Alightings} &
  {0.985*} &
  {0.984*} &
  {0.967} &
  {40.084$\times x$-271.740} \\
{} &
  {} &
  {Boardings} &
  {0.843*} &
  {0.919*} &
  {0.845} &
  {43.142$\times x$-13325.953} \\
{} &
  \multirow{-2}{*}{{Weeday}} &
  {Alightings} &
  {0.836*} &
  {0.908*} &
  {0.824} &
  {43.664$\times x$-12398.272} \\
{} &
  {} &
  {Boardings} &
  {0.888*} &
  {0.918*} &
  {0.842} &
  {42.847$\times x$-531.369} \\
\multirow{-4}{*}{{Station}} &
  \multirow{-2}{*}{{Hour}} &
  {Alightings} &
  {0.874*} &
  {0.907*} &
  {0.822} &
  {41.860$\times x$-451.766} \\ \hline
\multicolumn{4}{l}{* presents P-value less than 0.001.}
\end{tabular}
\end{table*}

\subsection{Data integration}

A random forest model \cite{Hastie2017} is used to integrate the APC ridership data with the crowd-sensing data from TravelSense. From the crowd-sensing data a feature matrix is constructed composed of the mode and OD data. For each station, the vectors corresponding to travel modes, origins, and destinations are concatenated into a single vector. The vectors for all stations then compose the feature matrix for the whole system. We use the random forest model to build a regression for the ridership based on the travel modes and OD structure associated with each station from the feature matrix. To obtain more insights for understanding the commuting patterns, we use two global model-agnostic methods: Permutation Feature Importance (PFI) \cite{Molnar2022} and the Partial Dependence Plot (PDP) \cite{Molnar2022}.

In the PFI a feature is considered important if in the random forest model it is frequently selected as a non-leaf node in the decision trees that make up the forest. Therefore, the importance of a feature can be quantified by the number of times that the feature is selected as a non-leaf node divided by the total number of non-leaf nodes.

PDP is used to measure the marginal effect of each feature on the ridership. This is done by changing the value of each studied feature in turn, while keeping the values of the other features fixed.




\section{Results}

\subsection{Representativeness of TravelSense}

\begin{figure}[]
\centerline
{\includegraphics[width=9cm]{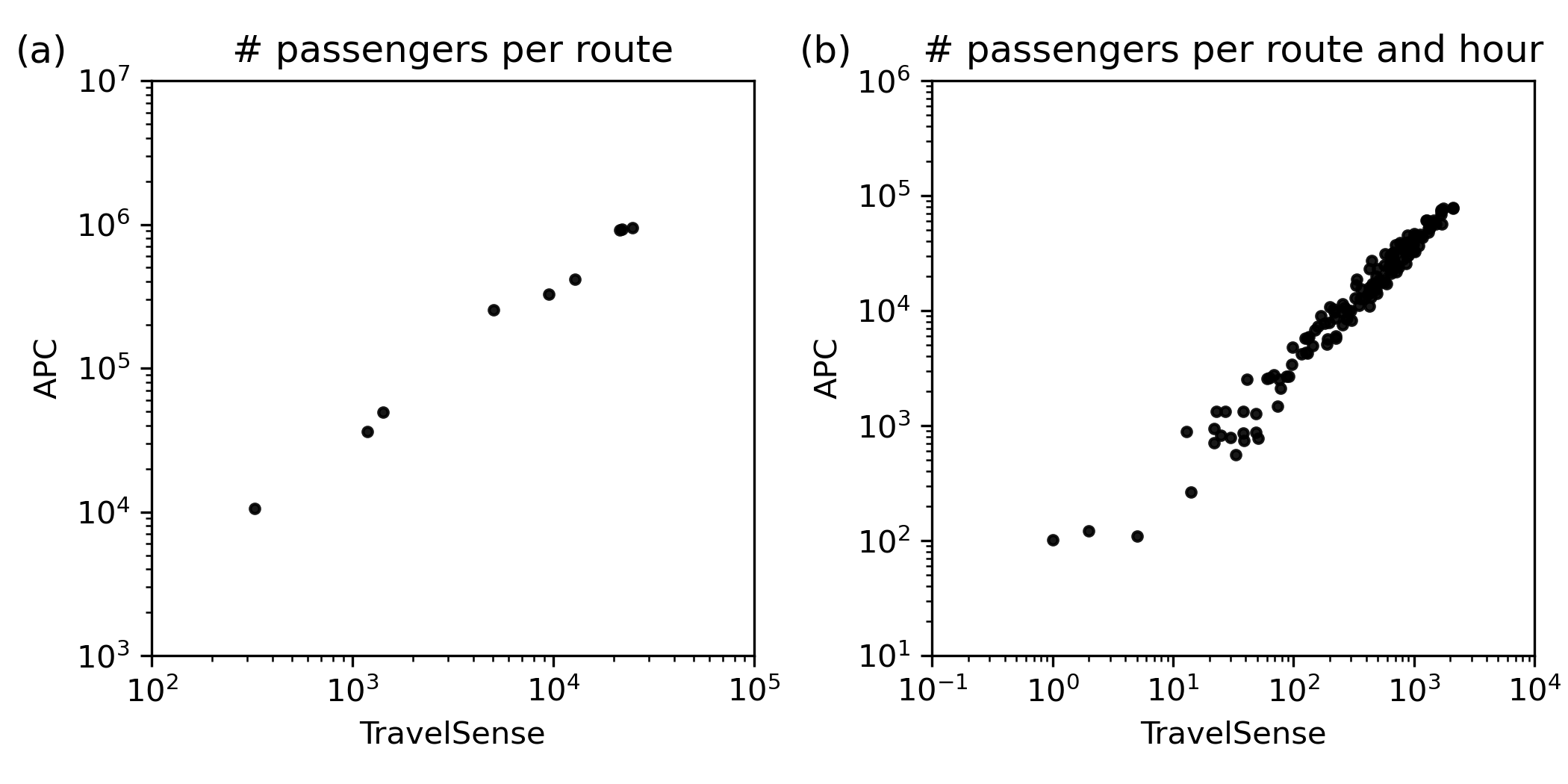}}
\caption{Comparison of boardings between TravelSense and APC considering (a) routes and (b) routes per hour.}
\label{fig:correlation_route}
\end{figure}

\paragraph{Route-level validations}  Fig.~\ref{fig:correlation_route}(a) compares APC and TravelSense boarding demand for nine train routes. The correlation found between the two data sources is 0.991, for Spearman's rank correlation, and 0.982 using the PCC value. This suggests that TravelSense is able to capture the route's ridership accurately. In Fig.~\ref{fig:correlation_route}(b) boarding counts are aggregated hourly per route; there is a data point per hour for each route. In this case Spearman's rank correlation is 0.985 and PCC is 0.984. Therefore TravelSense can also sense the temporal ridership trends for different routes.

\begin{figure}[tb]
\centerline
{\includegraphics[width=9cm]{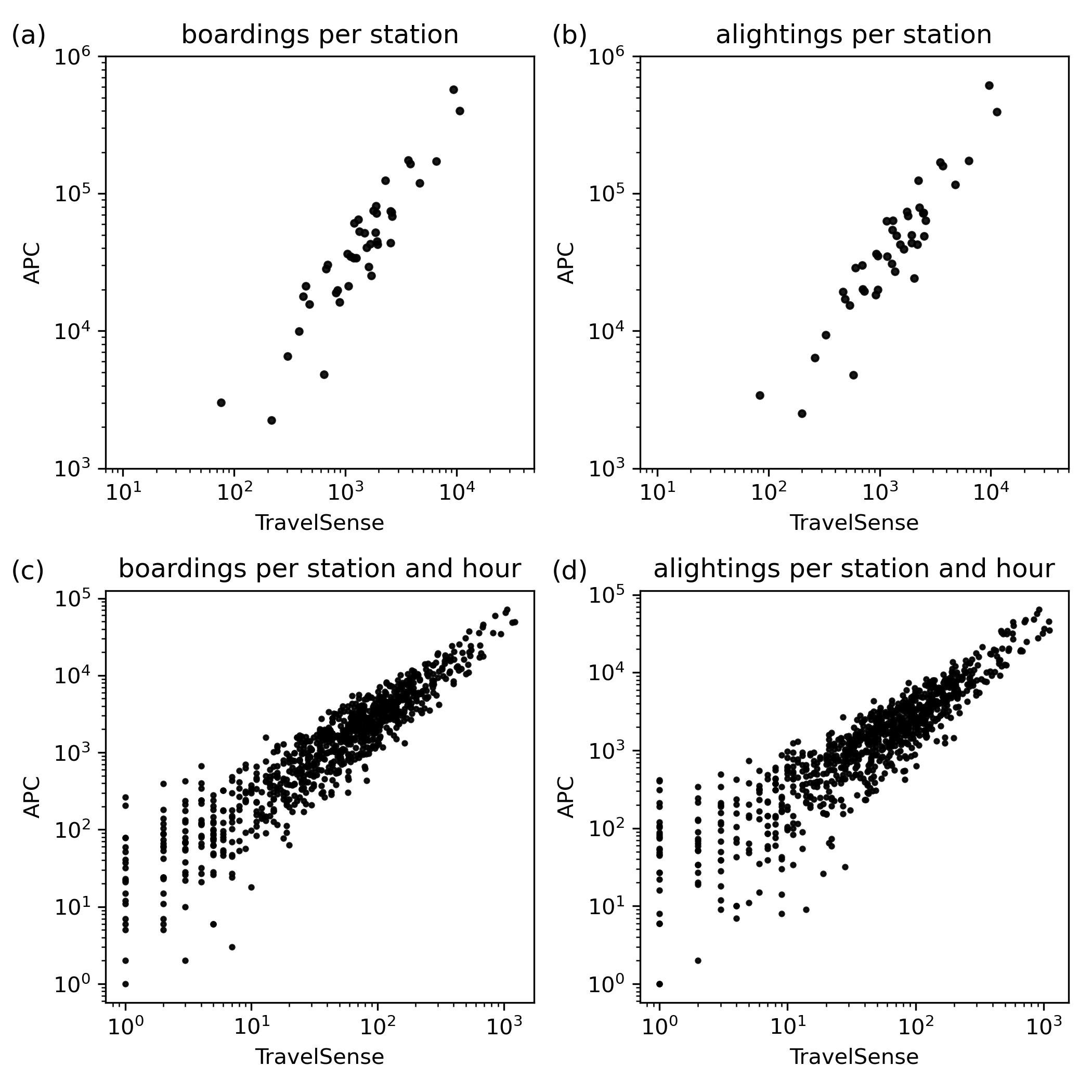}}
\caption{Comparison of station demand between TravelSense and APC considering (a) boardings per station, (b) alightings per station, (c) boardings per station and hour (d) alightings per station and hour.}
\label{fig:correlation_station}
\end{figure}

\begin{figure*}[htb]
\centerline
{\includegraphics[width=16cm]{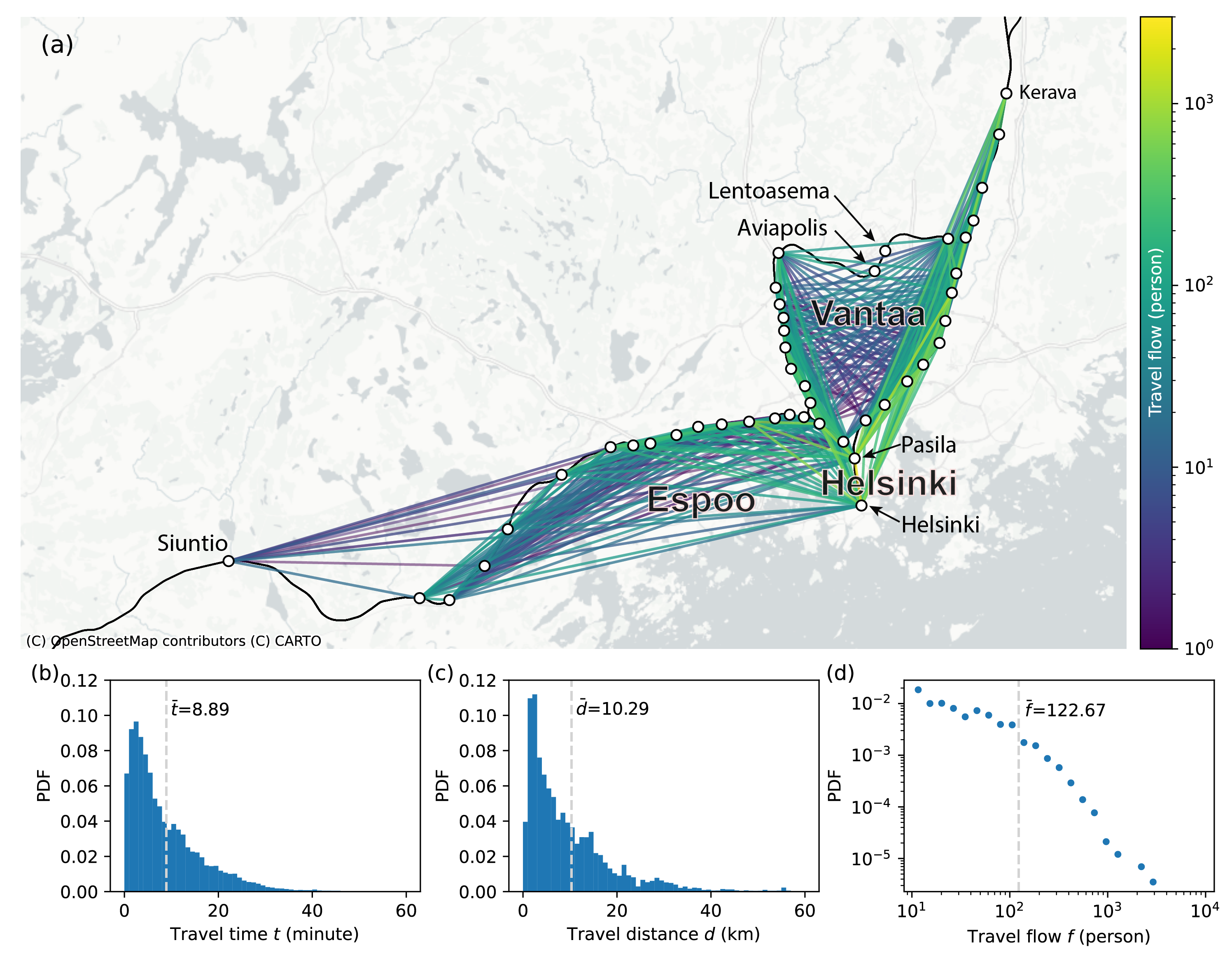}}
\caption{Commuting mobility patterns based on TravelSense. (a) visualization of the OD matrix; (b) travel time distribution; (c) travel distance distribution; (d) distribution of travel flows between OD pairs. In (b)-(d) the mean values are shown with dashed lines.}
\label{train_mobility}
\end{figure*}

\paragraph{Station-level validations} The demand at train stations is different than for the routes. To properly capture the passengers that use a station we consider boarding and alightings separately. In general the boarding and alighting numbers are different for the same train station, additionally some stations in Helsinki serve very large number of routes (e.g., Pasila and the Helsinki central station). As shown in Table.~\ref{table_correlations} and Fig.~\ref{fig:correlation_station}, with or without temporal factors, TravelSense both shows high positive correlations between APC station ridership --- Spearman's rank correlation ranges from 0.836 to 0.874 --- and is statistically significant (i.e., P-value is less than 0.001). The $R^2$ value of regression of hourly boardings is 0.842, while for hourly alightings it is 0.822. There is a potential in future works to scale the TravelSense data up according to its regression with APC ridership data; the scaling amplitude is around 43 (the coefficient of $x$, see Table.~\ref{table_correlations}) and gives an idea of the penetration rate of TravelSense sharing among rail passengers. However, it is noticed that in the raw data of TravelSense, records of two train stations (i.e., the station at the airport -- Lentoasema and an adjacent one -- Aviapolis) are missing. Therefore, the data quality could be improved and the level of improvement is yet to be explored.

\subsection{Commuting patterns within the commuter train system}

For the passengers that share their data, TravelSense makes it possible to identify the exact vehicles and routes that they use at the stations where they access and egress the rail network. This information allows identifying passenger flows, travel times, and travel distances which can not be done with APC alone.

Fig. \ref{train_mobility}(a) shows a visualization of the OD matrix constructed from TravelSense data. Generally, people living in the east of both Helsinki and Vantaa are more reliant on commuter trains than people who live in Espoo. The top two stations used are Helsinki railway station and Pasila which are geographically central and also serve most routes. Fig. \ref{train_mobility}(b) shows the travel time distribution for train legs. The average travel time is about nine minutes. According to HSL timetable data, the average travel time between two adjacent stations is about \textit{2.5} minutes which means, on average, a train leg consists of between three to four stops.
Fig. \ref{train_mobility}(c) shows the travel distance distribution for train legs. The average distance of commuter train trips is 10.29 km, the average speed of commuter trains is around 69.5 km/h yielding the average travel time of nine minutes.

Broadly, these results are in line with HSL's travel survey \cite{Raty2018}, which reports individual journeys across all travel modes took on average 24 minutes with an average journey length of 7.3 km. Therefore commuter train legs are shorter in time while covering larger distances than an average journey, as expected for the type of travel mode.

Fig. \ref{train_mobility}(d) shows the observed statistical distribution of travel flows between station pairs. The average travel flow sensed by TravelSense is 122 passengers and the maximum travel flow sensed by TravelSense is 2,735 passengers.

\subsection{Mobility beyond the train system}

\begin{figure*}[tb]
\centerline
{\includegraphics[width=18.5cm]{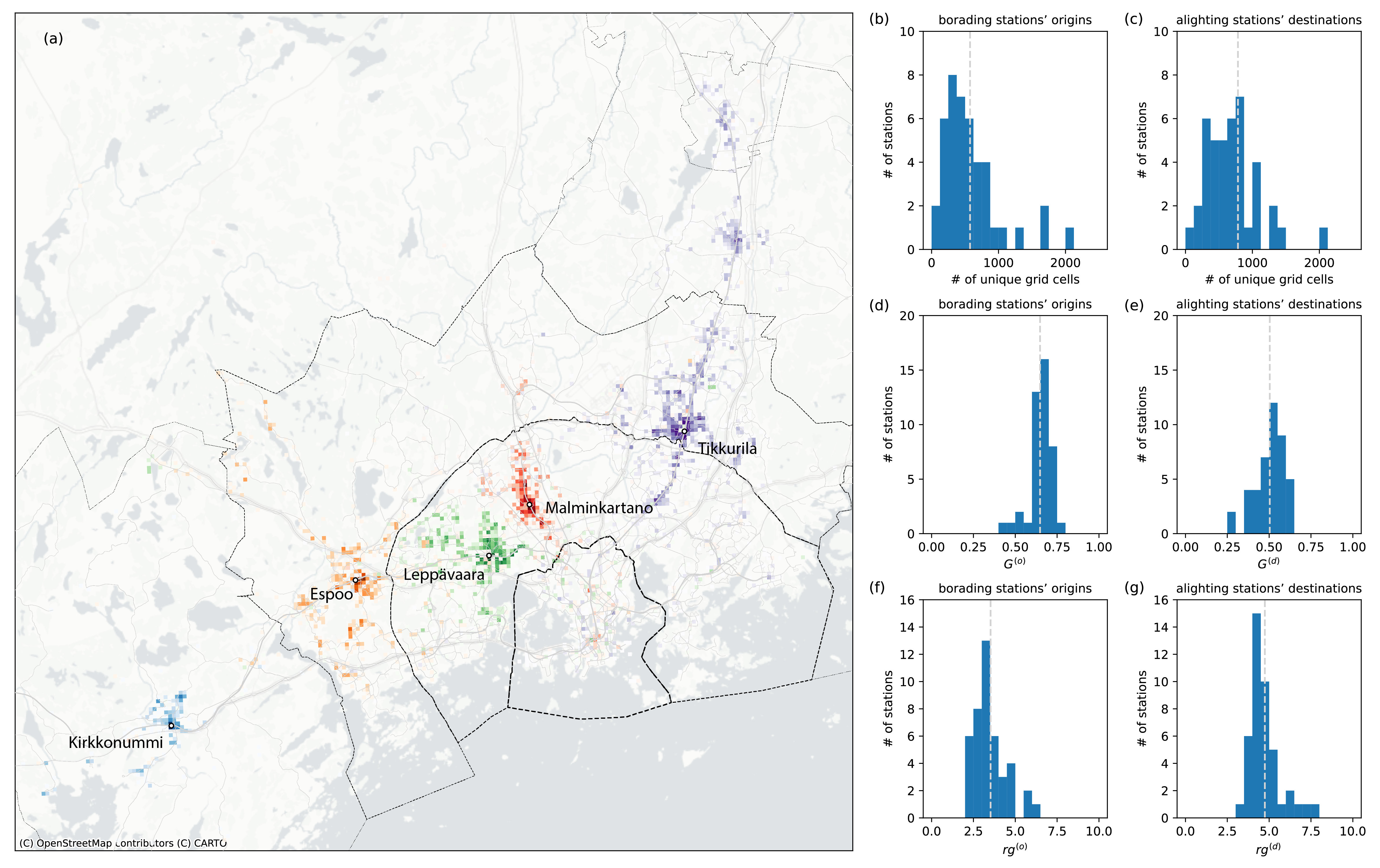}}
\caption{Distributions of boarding stations' origins and alighting stations' destinations. (a) Distributions of borading stations' origins for five stations. From left to right, Kirkkonummi (blue), Espoo (orange), Leppävaara (green), Malminkartano (red) and Tikkurila (purple). Gray solid lines represent the PT network, the saturation of a grid increases with the increment of passenger numbers of origins; (b) Number of origin grid cells; (c) Number of destination grid cells; (d) GINI index of origins; (e) GINI index of destinations; (f) Radius of gyration of origins; (g) Radius of gyration of destinations. Dashed lines in (b)-(g) show mean values.}
\label{fig:origins}
\end{figure*}

\paragraph{Origins and destinations for train stations} The commuter train system's ridership is influenced by the origin and destination of the trip chains (see Fig.~\ref{illustration}) of the passengers. 

By identifying the grid cells from which the passengers start and end their trip chains, the service coverage areas of stations can be determined. Fig.~\ref{fig:origins} (a) illustrates the distribution of boarding station's origins for five different train stations, which vary in number and geographical distribution.

The Tikkurila train station is a particularly noteworthy since its origins are not only located around the station, but also along the train line and bus lines, with several distant spots (Purple grid cells in Fig.~\ref{fig:origins} (a)). The reason for this is partly the location and function of this station in the PT network. It is located in the center of 4th most populous municipality, Vantaa, and also servea as a top-tier transfer hub \cite{Holm2016} in the HSL PT network design. It serves both local trains as well as longer distance routes that are also used by commuters coming into the capital region.

In order to quantify the patterns of origins and destinations, the three metrics described in section \ref{sec:OD_id} were applied to all train stations. As shown in the distributions of Fig.~\ref{fig:origins} (b) and (c), stations have more diverse destination than origins. Fig.~ \ref{fig:origins} (c) and (d) show that the passenger numbers distribution of origins is more concentrated than that of destinations. The average radius of gyration for destinations (Fig.~\ref{fig:origins} (e)) is larger than that of origins (Fig.~\ref{fig:origins} (f)),  indicating that when leaving the train station, passengers tend to travel to more remote locations. This asymmetry suggests that trains are potentially used earlier in the trip chain than other modes.

\begin{figure}[tb]
\centerline
{\includegraphics[width=9cm]{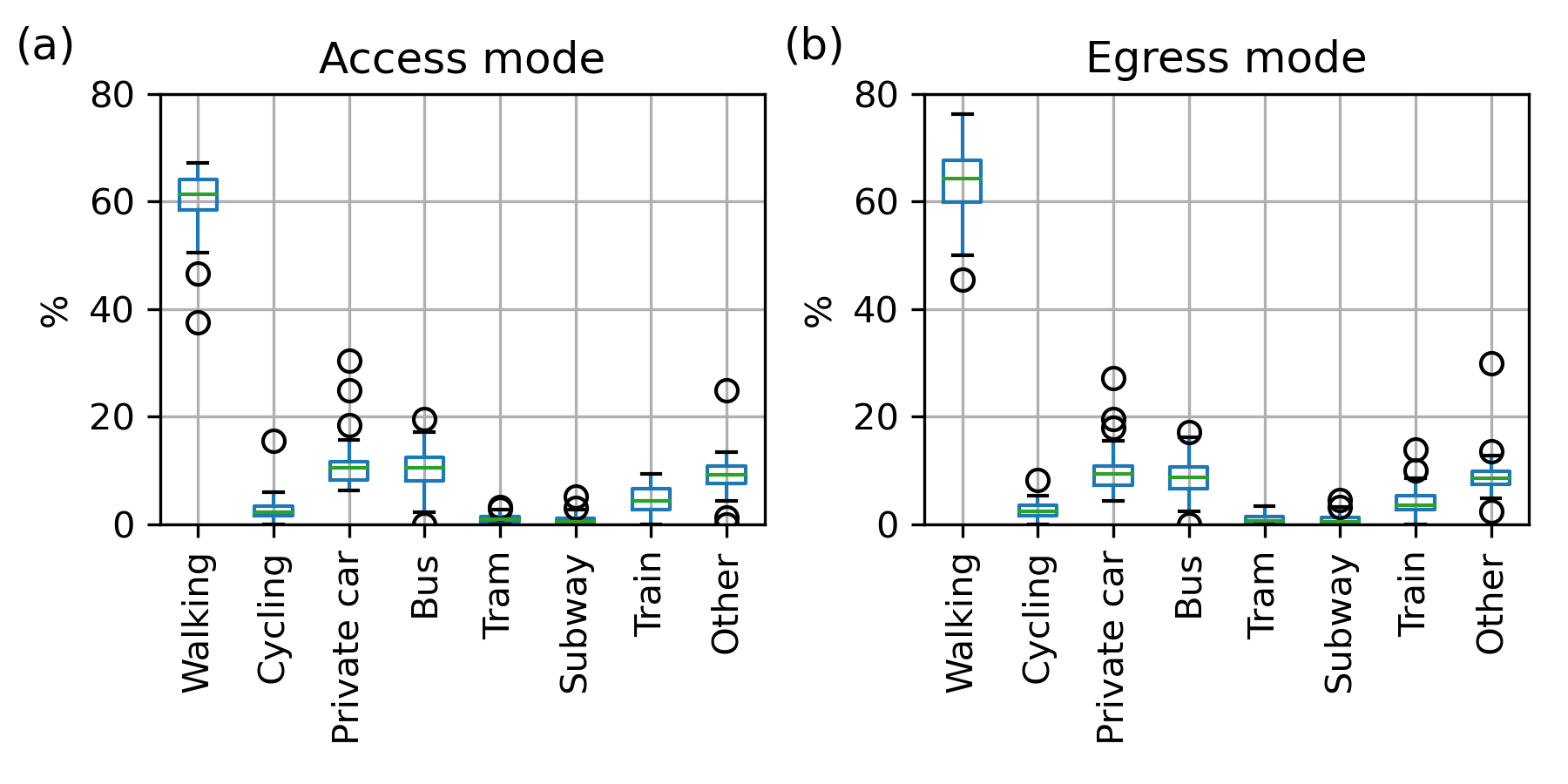}}
\caption{(a) Proportions for access modes and (b) proportions for egress modes.}
\label{fig:boxplot_modes}
\end{figure}

\paragraph{Traffic modes for access/egress to commuter train stations} The traffic modes used by passengers to access or leave a train station are crucial in designing a transfer-friendly PT system. As shown in Fig.~\ref{fig:boxplot_modes}, walking is by far the most commonly used mode of transportation to access stations. Private car and bus are also commonly used modes. Interestingly, they have similar proportions, with cars having a narrower distribution (although more outliers). The proportions of tram mode and subway mode are lower than 5\%, because only few train stations are easy to transfer to/from subways or trams by the current PT network structure. It is noted that modes designated as 'Other' in TravelSense data are either 'running', 'ferry' or not recognized.

\begin{figure}[tb]
\centerline
{\includegraphics[width=7.4cm]{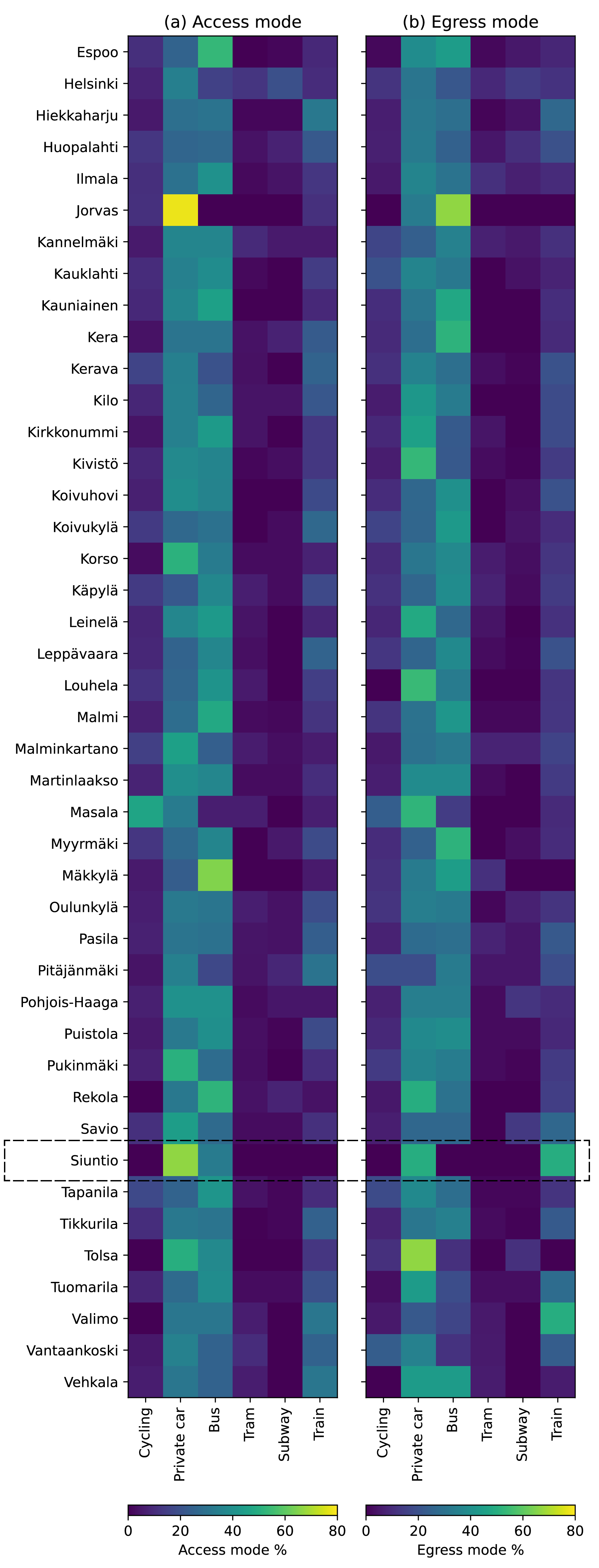}}
\caption{(a) Mode vector of access modes and (b) Mode vector of egress modes. Dotted line box shows the mode vector of Siuntio.}
\label{fig:modes}
\end{figure}

To further understand the mode choices of commuter train passengers, here, we focus on studying the six proportion of modes (cycling, private car, bus, tram, subway and train, see Fig.~\ref{fig:modes}). We use the Siuntio train station (labeled in Fig.~\ref{train_mobility}) as an example, where the main modes are bus and private vehicle, most likely due to its location at the westernmost part of the commuter train network, and more disperse residential areas. It has the third largest radius of gyration of origins ($rg^{(o)}_{\text{siuntio}}=5.87 > \bar{rg}^{(o)}=3.52$, the average) after Kerava and Tikkurila. In terms of egress modes (Fig.~\ref{fig:modes} (b)), the bus proportion is lower than for the access. There is also a larger use of trains, suggesting that some passengers are traveling beyond the Helsinki metropolitan area. 

\subsection{Data integration}

\begin{figure}[tb]
\centerline
{\includegraphics[width=7cm]{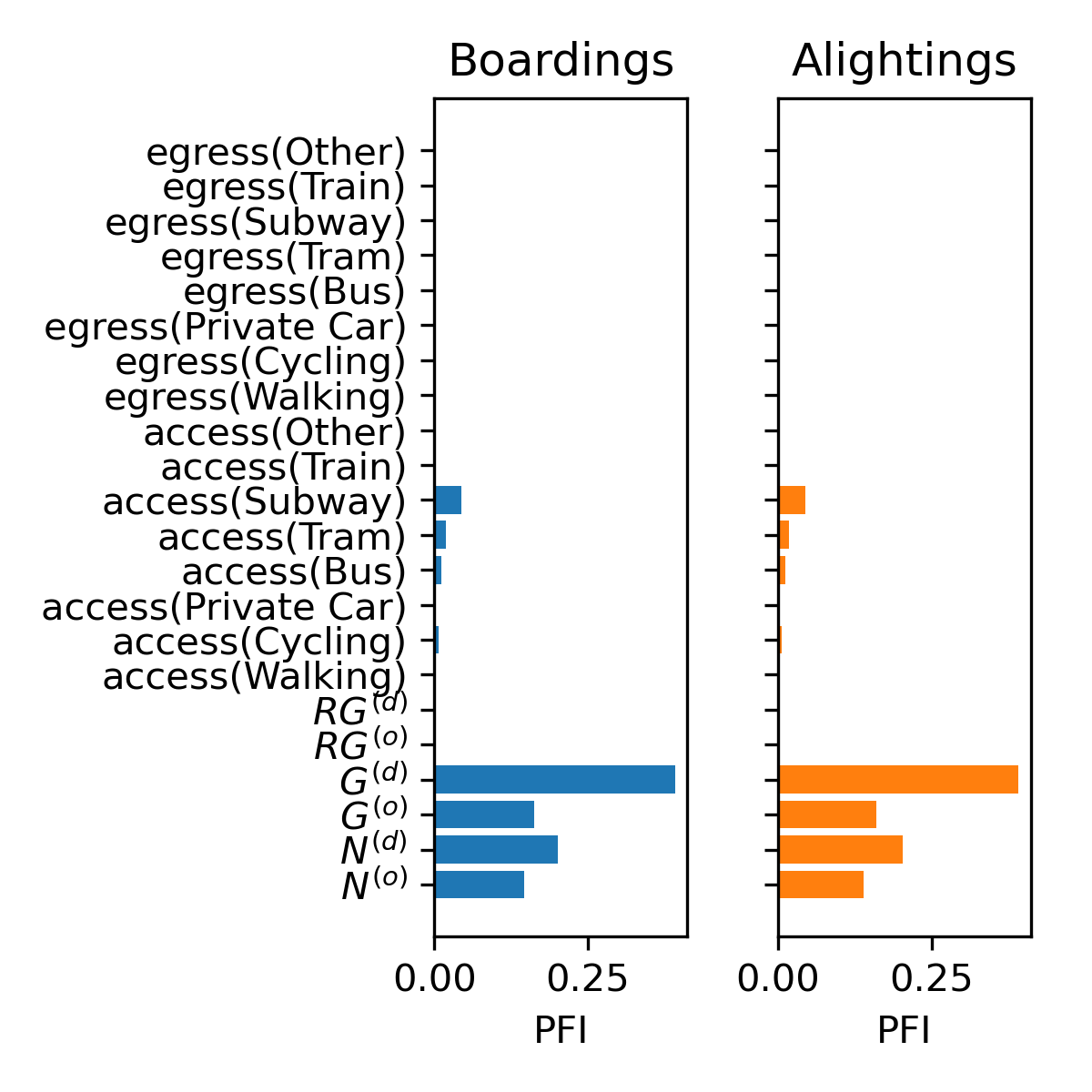}}
\caption{Permutation feature importance for commuting patterns and ridership.}
\label{fig:pfi}
\end{figure}

The data integration step combines two data sources via regression using a random forest model. The correlation given by the $R^2$ value for estimating the number of boardings is 0.935, and for the number of alightings, 0.934. Fig.~\ref{fig:pfi} shows that the contribution of the Gini index for destinations, $G^{(d)}$, is the most important feature for both the boardings and alightings. The second most important feature is the number of destinations $N^{(d)}$, followed by $G^{(o)}$ and $N^{(o)}$ for origins. Some access modes are also identified as having an impact via PFI. Namely subway, tram, bus and cycling, in order of imporatnce. This indicates that the destination of trip chains has a larger influence on passengers' choice of using trains (i.e., whether passengers are willing to use commuter train services or not). This is a hypothesis to be explored in future research.

We note that radius of gyration and egress modes are shown to play a minor role in the ridership model, which indicates that the diversity in popularity destinations and origins is more important than their spatial distribution.

\begin{figure}[tb]
\centerline
{\includegraphics[width=9cm]{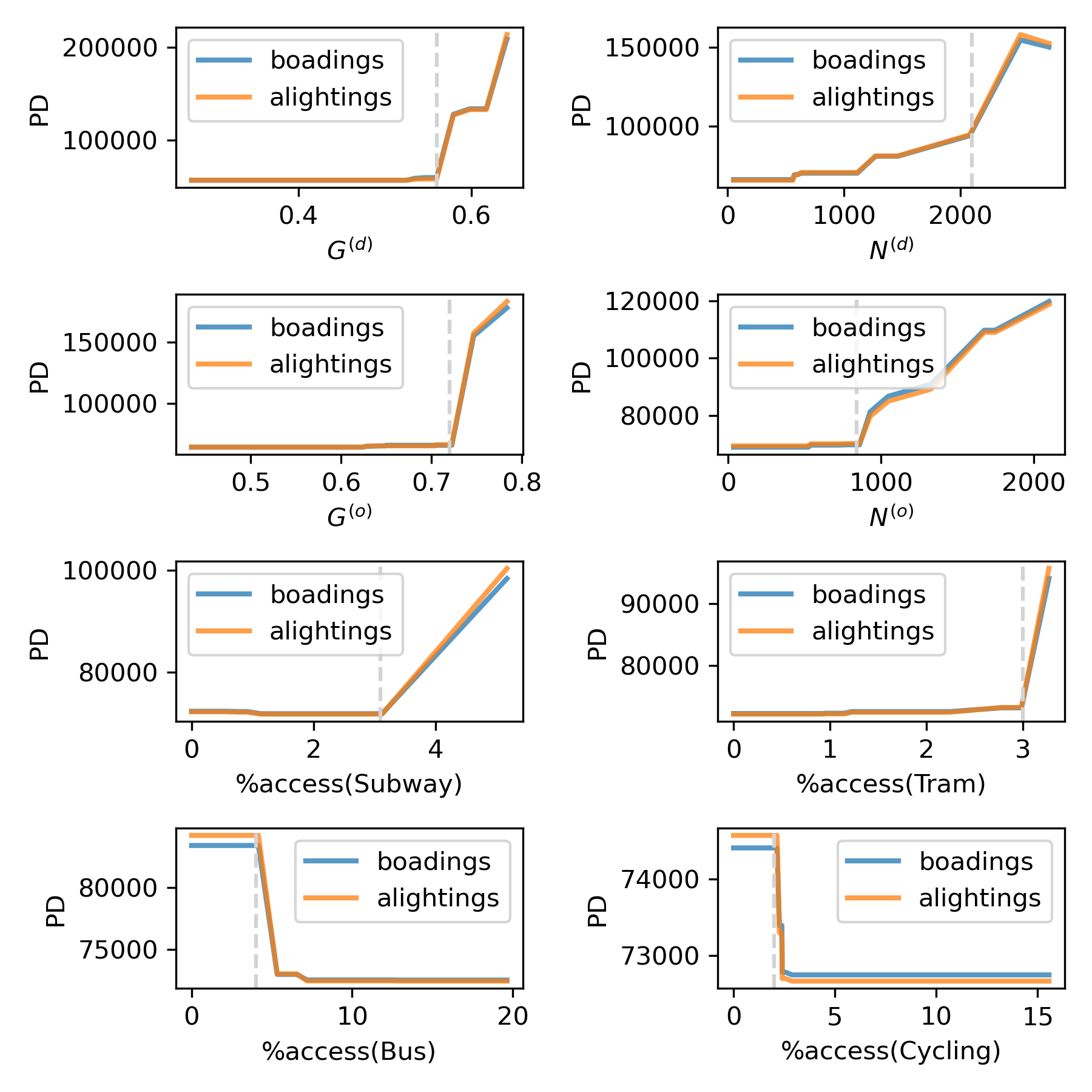}}
\caption{Partial dependence plots for commuting patterns and ridership.}
\label{fig:pdp}
\end{figure}

PDP is then used to explore how the top eight previously identified features affect the number of boardings and alightings. Fig. \ref{fig:pdp} shows the sensitivity of the predicted ridership to these features. We find that both alightings and baoardings behave similarly, they increase quickly with the destinations, $N^{(d)}$, in particular for values larger than $2100$ and with the number of origins when $N^{(o)}$ is larger than 840. As for the Gini coefficients, ridership shows little change until $G^{(d)}$ is larger than 0.56, from then on it rises sharply in steps. Similarly for $G^{(o)}$ the threshold value is around 0.72, with a sharp increase afterwards and a less marked step behaviour.

The importance of the features is reflected by the nonlinear relationships to the ridership, the travel mode features show a type of activation/deactivation, while for the OD features the dependency exhibits more complex behaviour. 

\section{Implications of practice}

APC devices are a common technology for counting the passengers entering and exiting PT vehicles. However, they are often not installed in all vehicles of a specific mode or route. Technical problems are common which can lead to gaps in measurements. An additional limitation is the aggregated nature of the information they provide which does not allow detailed analyses nor further insights beyond the PT network. 

Data from mobile devices, on the other hand, are more promising, however, their low maturity and penetration within PT networks presents a challenging task. The main challenge is the low volumes of data requiring validation and scaling in order to be trusted. Hence, their combination with more conventional data sources is important, if not mandatory, for the early stages of their implementation.  

The results of our analysis of the TravelSense dataset show that despite its limitations related with low volume of information, it can indeed be scaled up using data from conventional APC devices installed in all commuter trains within the Helsinki network. While more thorough validation is necessary (see \cite{Huang2022}), it is shown to be representative.

The representativeness of TravelSense data allows it to be useful, either alone or in combination with other data, for several applications. For example, TravelSense can be scaled using historical data to fill the gaps in APC measurements when APC devices have technical problems that hinder their operation \cite{Roncoli2023}. In addition, the representativeness of TravelSense dataset allows for estimation of passenger flows within the train network, and more specifically, the probabilities of a passenger travelling to any station based on the boarding station. Multiplying the probabilities from TravelSense data within a given time period with the APC historical counts can give estimated flows.

TravelSense was found to be useful for offering insights on the movements of passengers outside the train network. Such a notion can be utilized for a plethora of applications within the transport sector, for instance in studies related with multimodality and accessibility \cite{Kujala2018, Weckstrom2021}. Which would allow for enhancing the understanding of travelers' habits and their connection to PT systems.

It is apparent from this study that mobile data are associated with certain opportunities for PT systems, as long as there is enough background for filling gaps they might have as a result of their early introduction into the mobility sector. This study offers several numerical outputs related with the train mobility within Helsinki, but also insights on how such a dataset can be utilized within PT and more specifically for commuter train related applications. However, it is noted that the nature of this dataset allows it to have several multimodal applications, if properly processed and fused with additional information and data sources \cite{Zheng2015, Huang2018, Weckstrom2019, Wang2022, Qiu2022}. 




\section{Conclusion}
This study focused on using multi-source wireless data, such as APC and the TravelSense mobile phone application, in order to understand the commuting patterns in the Helsinki metropolitan area. The representativeness of ridership data was validated using both data sources, showing the ability of TravelSense to sense near-real-time PT states. The combination of the data sources provides in-depth information about origin and destination locations, traffic modes used to access/egress commuter train stations, and number of passengers boarding and alighting at each station. The proposed data integration framework could be used to inform PT planning decisions and promote sustainable transportation options. This study highlights the potential of the multi-source wireless data approach for efficient transportation management and operations with an emphasis on smooth multimodality and sustainability.

Addressing several challenges associated with such dataset and the presented integration framework (e.g., generating commuting features manually) by means of different approaches (e.g., by considering applying cross-domain data fusion methods, graph neural network models and transfer learning methods) is part of future work. The potentials of fusing or integrating this dataset with additional conventional PT data sources (e.g., smartcard data and General Transit Feed Specification (GTFS) data) are also envisioned extensions of this study.

\section*{Acknowledgment}

The authors thank HSL for access to the TravelSense dataset, and for their time and discussions about this study, We especially thank Pekka Räty and Hanna Kitti. The work of Z. Huang was supported by the Strategic Research Council at the Academy of Finland (grant numbers 345188 and 345183). The work of C. Sipetas was supported by the FinEst Twins Center of Excellence (H2020 Grant 856602). The work of A. Espinosa Mireles de Villafranca was supported by the Academy of Finland (grant: Profi4). Calculations were performed using computer resources within the Aalto University School of Science “Science-IT” project.

\vspace{12pt}

\end{document}